\documentclass{Interspeech2024}
\interspeechcameraready
\usepackage{xcolor}
\usepackage{manyfoot}

\usepackage[utf8]{inputenc}
\title{LiteFocus: Accelerated Diffusion Inference for Long Audio Synthesis}

\name[affiliation={}]{Zhenxiong}{Tan}
\name[affiliation={}]{Xinyin}{Ma}
\name[affiliation={}]{Gongfan}{Fang}
\name[affiliation={*}]{Xinchao}{Wang}

\address{
  National University of Singapore, Singapore
}
\email{zhenxiong@u.nus.edu, maxinyin@u.nus.edu, gongfan@u.nus.edu, xinchao@nus.edu.sg}


\newcommand{\methodname}{LiteFocus}
\newcommand{\efficiency}{1.99}

\begin{document}

\maketitle

\begin{abstract}

    Latent diffusion models have shown promising results in audio generation, making notable advancements over traditional methods. 
However, their performance, while impressive with short audio clips, faces challenges when extended to longer audio sequences. 
These challenges are due to model's self-attention mechanism and training predominantly on 10-second clips, which complicates the extension to longer audio without adaptation.
In response to these issues, we introduce a novel approach, \methodname\, that enhances the inference of existing audio latent diffusion models in long audio synthesis.
Observed the attention pattern in self-attention, we employ a dual sparse form for attention calculation, designated as \textit{same-frequency focus} and \textit{cross-frequency compensation}, which curtails the attention computation under same-frequency constraints, while enhancing audio quality through cross-frequency refillment.
\methodname\ demonstrates substantial reduction on inference time with diffusion-based TTA model by {\efficiency$\times$} in synthesizing 80-second audio clips while also obtaining improved audio quality.

\end{abstract}
{
\renewcommand{\thefootnote}{}
\footnote{\textsuperscript{*} Corresponding Author.}

}
\setcounter{footnote}{0}

\section{Introduction}

Text-to-audio (TTA) has become an increasingly important area of research, with practical applications that span speech synthesis~\cite{taylor2009text, wang2023neural}, music production~\cite{dong2018musegan, copet2024simple} and assistive technologies~\cite{sharma2012speech}. 
The recent progress in TTA has been significantly propelled by advancements in deep learning and the scaling up of models~\cite{yang2023diffsound, liu2023audioldm, liu2023audioldm2, kreuk2022audiogen, dong2018musegan}.
Among them, the application of latent diffusion models~\cite{rombach2022high}, originally developed for image and video generation,  shows a significant leap forward to audio synthesis~\cite{yang2023diffsound,liu2023audioldm,liu2023audioldm2}.
This has led to notable improvements in audio fidelity, showcasing the latent diffusion model's capacity to elevate the quality and feasibility of audio content creation.

Despite the significant successes of diffusion-based model in audio synthesis, the model encounters efficiency challenges.
Previous work on accelerating diffusion-based TTA models focuses on reducing the timesteps by progressive distillation~\cite{huang2022prodiff,mehta2023matcha} or consistency distillation~\cite{ye2023comospeech,bai2023accelerating}. 
Another line of work is to accelerate the sampling of diffusion by SDE~\cite{vovk2022fast} or ODE~\cite{chen2023lightgrad,guo2023voiceflow} sampling methods. 
Those methods all successfully reduce the sampling steps and thus increase the efficiency. 

\begin{figure}
    \centering
    \includegraphics[width=0.5\textwidth]{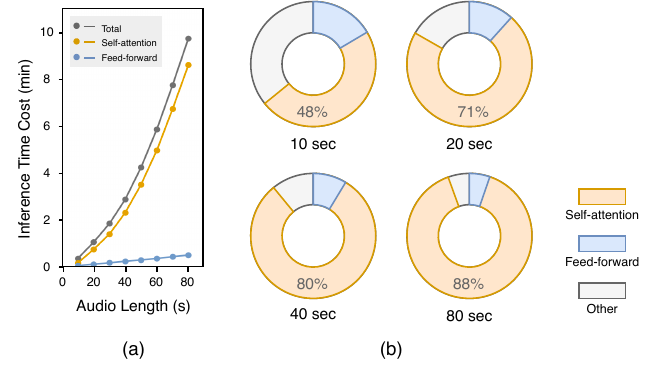}
    \caption{Illustration of the computational cost associated with audio synthesis using the audio diffusion model (AudioLDM2): (a) The time required for synthesizing audio clips of varying lengths; (b) The proportional time consumption of different modules within the model during the generation of audio clips of different lengths.}
    \label{fig:result}
\end{figure}

However, beyond reducing the number of inference steps for diffusion models, it is necessary to place emphasis on special time-consuming structures within the diffusion model. 
A primary issue is the drastic increase in inference time as the length of the generated audio extends~\cite{evans2024fast}, as the self-attention mechanism~\cite{vaswani2017attention} within the unet model~\cite{ronneberger2015u} has $O(N^2)$ complexity. 
For instance, generating an 80-second audio on the model \textit{AudioLDM2} takes approximately 10 minutes (details in Section~\ref{sbs:setup}).
In Figure \ref{fig:result}, we show that the inference time exhibits a quadratic increase with the length of the audio. Furthermore, as the length of the audio increases, self-attention gradually dominates the overall inference time.
This increasing proportion underscores the importance of enhancing the efficiency of self-attention calculation in audio diffusion models under the scenario of long-form audio synthesis.

Inspired by this, we focus on the acceleration of attention mechanism in diffusion-based TTA model. Our approach does not require retraining the model, making it cost-effective and efficient.
Our contributions can be summarized as follows: 
\begin{itemize}
    \item We uncover a distinctive attention pattern specific to the mel-spectrogram within the attention blocks. The high interactions within the same frequency band reveals the redundancy of attention computation. 
    \item Drawing on these findings, we propose \textbf{\methodname}, a method that sparisifies the computation of attention. We factorize the computation of attention into two parts: sample-frequency focus and cross-frequency compensation.
    \item Our experimental evaluations suggest that \methodname\ not only potentially reduces inference time of audio diffusion models but also may enhance the quality for longer audio segments in comparison to baseline methods. 
\end{itemize}
\section{Related Works}

\noindent\textbf{Audio Diffusion Models.} Following the success in computer vision~\cite{rombach2022high,saharia2022photorealistic, ruiz2023dreambooth}, diffusion models have been adeptly adapted for tasks such as text-to-speech synthesis and speech enhancement, demonstrating remarkable capabilities~\cite{jeong2021diff, kong2020diffwave, yang2023diffsound, liu2023audioldm, liu2023audioldm2}.
Earlier models like \textit{DiffWave}~\cite{kong2020diffwave} and \textit{Diff-TTS}~\cite{jeong2021diff} showcased diffusion models' potential in audio synthesis. 
Recently, advancements with Diff-sound~\cite{kong2020diffwave} and \textit{AudioLDM}~\cite{liu2023audioldm,liu2023audioldm2} leveraged larger datasets and models to achieve superior audio quality, illustrating diffusion models' growing capability in high-quality audio generation from text.

\noindent\textbf{Accelerating Audio Synthesis Model.}
To speed up audio synthesis, methods such as \textit{InferGrad}~\cite{chen2022infergrad} use joint training to make inference faster, \textit{ProDiff}~\cite{huang2022prodiff} applies knowledge distillation to cut down on the number of steps needed, and techniques like \textit{Diffsound}~\cite{yang2023diffsound} and \textit{NoreSpeech}~\cite{yang2022norespeech} use VQ-VAE~\cite{van2017neural} for quicker token generation. 
Additionally, some general methods for speeding up diffusion models~\cite{bolya2023token, yang2023diffusion, fang2024structural, ma2023deepcache} can also be applied in audio synthesis.
For instance, \textit{Token Merging}~\cite{bolya2023token} optimizes the transformer module, offering a direct approach to enhancing the efficiency of diffusion models. 
\section{Methods}
\subsection{Preliminary}
The audio latent diffusion model~\cite{yang2023diffsound, liu2023audioldm,liu2023audioldm2} can  be succinctly divided into three main components: the text encoder~\cite{elizalde2023clap}, the denoise U-Net model~\cite{ronneberger2015u}, and a mel-spectrogram-based variational auto-encoder (VAE)~\cite{kingma2013auto}. 
The pipeline for converting text to audio operates as follows: 
Initially, an input text is processed by the text encoder, which encodes the text into a corresponding embedding. 
Subsequently, this specific text embedding serves to guide the iterative denoising of a sampled noise within the latent space, executed through a U-Net architecture.
Upon completion of the denoising process, the latent representation is decoded by the VAE decoder into a mel-spectrogram, which can then be converted into audible sound.

Within the inference process of the audio latent diffusion model, self-attention modules play a pivotal role as one of the key units.
The self-attention module takes an input \(X \in \mathbb{R}^{(N_t \times N_f) \times C}\) and transforms it into corresponding keys, queries, and values.
 \(C\) stands for the number of channels, and \(N_t\) and \(N_f\) delineate the dimensions within the latent space, corresponding respectively to the time and frequency bands of the mel spectrogram. 
The fundamental formula for computing self-attention can be expressed as follows: 
\begin{align*}
    &Y = \text{attention}(X) = \text{softmax}\left(\frac{QK^T}{\sqrt{d_k}}\right)V, \\
    &\text{where}\ K = W_kX, Q = W_qX, V = W_vX
\end{align*}
where \(d_k\) represents the dimensionality of the keys, and \(W_q\), \(W_k\) and \(W_b\) are the projection matrices.
This self-attention mechanism allows the model to weigh the importance of different parts of the input in relation to each other, enhancing its ability to generate coherent and contextually relevant audio outputs. 
However, in the context of audio latent diffusion, the computational load of the $QK^T$ exhibits quadratic growth w.r.t the length of the audio. 


\subsection{Attention Pattern for Audio Latent Diffusion}
\begin{figure}[h]
    \centering
    \includegraphics[width=0.45\textwidth]{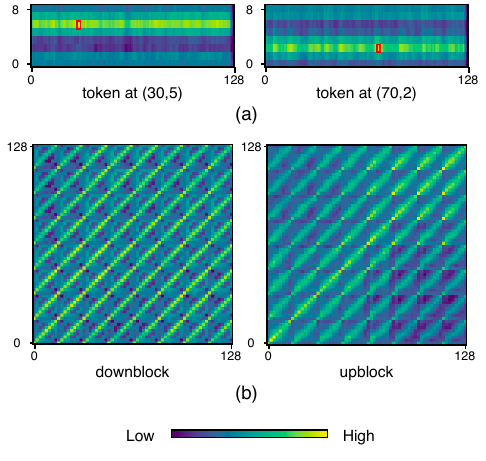}
    \caption{Attention patterns within the audio latent diffusion model: 
    (a) Attentions for two different tokens. We reshape the attentions to rearrange the attention of the same frequency in the same row. 
    (b) Attention patterns across different blocks in the model. Due to size constraints, only a proportion of the attention maps are shown.
    }
    \label{fig:att_pattern}
\end{figure}
We show the attention maps in Figure \ref{fig:att_pattern}.
The results highlight a unique attention pattern, focusing on interactions within the same frequency band. 
We observe that the token's attention is specifically directed towards other tokens sharing the same frequency coordinate. 
From Figure \ref{fig:att_pattern}(a), we observe that tokens of the same frequency have higher attention values relative to each other. 
In Figure \ref{fig:att_pattern}(b), this is shown as the attention pattern exhibiting equidistant repeating pattern, and the interval of this repetition corresponds to the number of frequency dimensions in the mel spectrogram within the latent space.
Additionally, we find that this pattern is more pronounced in the down-sampling block compared to the up-sampling block.

\subsection{\methodname}

\begin{figure*}[ht]
    \centering
    \includegraphics[width=\textwidth]{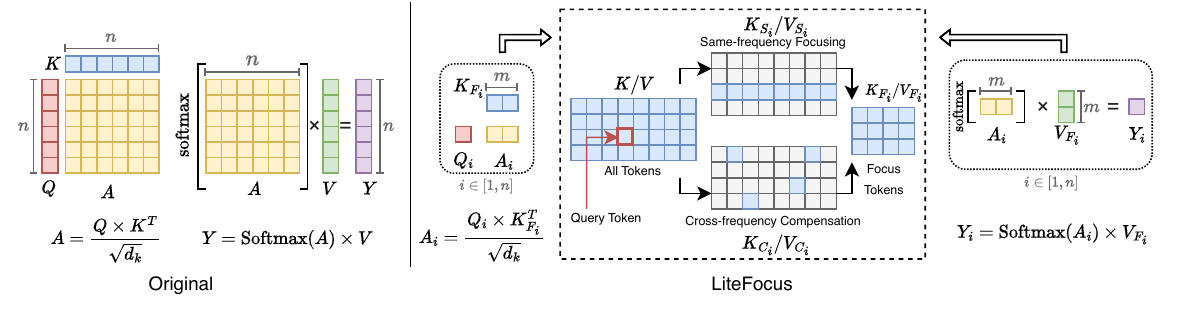}
    \caption{A comparison between the original attention processor and \methodname: The left side illustrates the processor for the original self-attention mechanism, while the right side depicts the attention processor utilized by \methodname. In \methodname\ , each query is assigned a specific focus tokens set, with which it performs the attention operation on the corresponding keys and values.}
    \vspace{-1em}
    \label{fig:liteFocus}
\end{figure*}





Motivated by the aforementioned approach, we posit that dedicating substantial computational resources to the repetitive patterns represents a promising avenue for optimization. 
Particularly for longer audio sequences, enhancing the efficiency of attention computations and eliminating superfluous operations stand to offer considerable improvements in inference speed, all while maintaining the quality of the generated audio.

To facilitate the streaming of audio diffusion model without extra post-training, we propose \methodname\ that sparsifies the attention mechanism.
Our approach focuses on the attention process by sparsifying each query's attention operation on a subset of all tokens, rather than computing attention across all keys and values for every token.
We factorize the sparsification of attention into two directions: Same-frequency Focusing and Cross-frequency Compensation.

To elaborate, within the attention mechanism, a token located in the mel-spectrogram latent space at coordinates \((a, b)\) is denoted as \(X_i\), where \(i\) is the index after flattening the dimensions, computed as \(i = a \cdot N_t + b\). 
Here, \(N_t\) represents the total number of time steps in the latent space, and \((a, b)\) specifies the token's position with \(a\) and \(b\) indicating its time and frequency coordinates, respectively.
For this token $X_i$, there is a specifically associated focus-set \(F_{i}\). 
This focus-set \(F_{i}\) represents the collection of indices from keys \(K\) and values \(V\) with which query \(Q_{i}\) is intended to interact. 
The output \(y_{i}\) obtained from this query is formulated as follows:
\begin{equation}
y_{i} = \text{softmax}\left( \frac{Q_{i}K_{F_{i}}^T}{\sqrt{d_k}} \right)V_{F_{i}}
\end{equation}
where $K_{F_{i}}$ and $V_{F_{i}}$  denote the subsets of keys and values selected for interaction  with $Q_{i}$, defined as:
\begin{align}
    K_{F_{i}} &= \left\{K_{j} | j \in F_{i}\right\}\\
    V_{F_{i}} &= \left\{V_{j} | j \in F_{i}\right\}
\end{align}
The focus-set \(F_{i}\) for each query \(q_{i}\) is formed by the union of two distinct sets:
\begin{equation}
F_{i} = S_{i} \cup C.
\end{equation}

\begin{itemize}
    \item Focus on Same-frequency Tokens.
    
    The set of same-frequency tokens \(S_{i}\) is composed of indices corresponding to tokens that are in the same frequency band as \(Q_{i}\).
    This can be formulated as:
    \begin{equation}
    S_{i} = \{j \mid j \bmod N_t = i \bmod N_t\}
    \end{equation}

    \item Compensation on Cross-frequency Tokens.

    The set of Cross-frequency Compensation \(C\) is obtained by randomly selecting indices from the full index set \(I\) of keys and values according to the percentage \(r\):
    \begin{equation}
    C = \text{RandomSample}\left(I, \lfloor r \cdot |I| \rfloor\right),
    \end{equation}
    where \(r\) represents the percentage of the total index set \(I\) to be included in the subset \(C\), and \(|I|\) denotes the cardinality (or total number) of \(I\). This selection process provides a global context through a diverse sample from \(I\), proportional to \(r\).
    
\end{itemize}
Figure~\ref{fig:liteFocus} illustrates the process of selecting the focus-set by \methodname\. 
Combining both the Same-frequency Focusing $S_i$ and the Cross-frequency Compensation $C$ allows the attention mechanism to achieve both a broad contextual understanding and a detailed insight into frequency-specific relationships within the audio.
By ensuring that each query interacts with only a limited number of tokens, our method effectively reduces the total computational load.

\section{Experiments}
\subsection{Setup}
\label{sbs:setup}
\textbf{Base Model and Inference Setting:}
In our experiments, we focused on the AudioLDM2 model, one of the standout models in the audio latent diffusion category, specifically using the \texttt{audioldm2-full} checkpoint~\footnote{
AudioLDM2: \href{https://github.com/haoheliu/AudioLDM2}{https://github.com/haoheliu/AudioLDM2}
}. 
The diffusion steps were set to the default 200 steps. 

\textbf{Evaluation:}
Our experiments utilized the AudioCap Evaluation Set~\cite{kim2019audiocaps}, generating 4845 unique audio clips from 4845 captions to serve as text prompts.
The quality and efficiency of these audio generations were evaluated using several metrics, including Frechet Audio Distance (FAD)~\cite{kilgour2018fr}, Kullback-Leibler (KL) divergence, Contrastive Language-Audio Pretraining (CLAP) score~\cite{laionclap2023} and the inference time required for each audio clip. 
The code used for these evaluation experiments is based on the repository \textit{Audio Generation Evaluation}\footnote{
Evaluation Tools: \href{https://github.com/haoheliu/audioldm\_eval}{https://github.com/haoheliu/audioldm\_eval}
}. 

\textbf{Infrastructure}
To ensure consistency and reliability in our measurements, especially for those related to computational costs during inference, all experiments were conducted on a single Nvidia A6000 GPU.

\textbf{\methodname\ Setting:}
For \methodname\, we applied it to all Transformer Modules within the second block of the down blocks and the second block of the up blocks of the audio latent diffusion model.
Furthermore, the proportion $r$ for Cross-frequency Compensation is set to 0.1.

\textbf{Baseline:}
We also applied the Token Merging~\cite{bolya2023token} approach to audio latent diffusion model as our baseline. 
Similar to our method, Token Merging aims to reduce the redundant computations within the attention operation.
Its primary strategy involves merging similar tokens into a single token before executing the attention operation, thereby reducing the computational load of the attention mechanism. 
Token Merging was applied to the same transformer modules as \methodname\ and utilized all default parameters.


\begin{table*}[h]
\centering
    \small
    \resizebox{\linewidth}{!}{
    \begin{tabular}{l  |c c c c  |c c c c|c c c c}
      \toprule
      & \multicolumn{4}{c|}{\bf Original Inference} & \multicolumn{4}{c|}{\bf \methodname\ } & \multicolumn{4}{c}{\bf Token Merging}\\
      \midrule
      \bf Audio Length& \bf FAD $\downarrow$& \bf KL $\downarrow$& \bf CLAP(\%)$\uparrow$& \bf Speed $\uparrow$ & \bf FAD $\downarrow$& \bf KL $\downarrow$& \bf CLAP(\%)$\uparrow$&  \bf Speed $\uparrow$ & \bf FAD $\downarrow$& \bf KL $\downarrow$& \bf CLAP(\%)$\uparrow$&\bf Speed $\uparrow$ \\
      \midrule
    10 sec& \textbf{2.89}& \textbf{1.82}& \textbf{19.8}& 1$\times$& 4.17& 1.86& 18.1& \textbf{1.02$\times$}& 3.73& 1.85& 19.7&1.01$\times$\\
      20 sec&  4.97& 2.01& 16.0& 1$\times$& \textbf{3.72}& \textbf{1.79}& \textbf{17.8}&   \textbf{1.43$\times$}& 7.67& 2.22& 15.5&1.25$\times$\\
      40 sec&  6.72& 2.27& 12.8& 1$\times$&  \textbf{5.40}& \textbf{2.12}& \textbf{13.8}&  \textbf{1.94$\times$}& 7.85& 2.43& 11.9&1.40$\times$\\
      80 sec& 7.74& 2.41& 10.8& 1$\times$& \textbf{6.56}&  \textbf{2.28}& \textbf{12.2}&  \textbf{1.99$\times$}& 8.95& 2.54& 10.2&1.56$\times$\\
      \bottomrule
    \end{tabular}
    }
    \caption{Efficiency and quality metrics of the original inference, \methodname, and Token Merging methods across different audio lengths. Bold values represent the best results among the three groups. In \methodname, the cross-frequency compensation ratio $r$ is set to 0.1.}
    \vspace{-1em}
    \label{tbl:main_stable_diffusion}
\end{table*}


\begin{table*}[ht]
\centering
    \small
    \resizebox{\linewidth}{!}{
    \begin{tabular}{l  |c c c   |c c c |c c c|cc c}
      \toprule
      & \multicolumn{3}{c|}{\bf 10 sec} & \multicolumn{3}{c|}{\bf 20 sec} & \multicolumn{3}{c|}{\bf 30 sec} & \multicolumn{3}{c}{\bf 40 sec}\\
      \midrule
      \bf $r$& \bf FAD $\downarrow$& \bf KL $\downarrow$& \bf CLAP(\%)$\uparrow$& \bf FAD $\downarrow$& \bf KL $\downarrow$& \bf CLAP(\%)$\uparrow$& \bf FAD $\downarrow$& \bf KL $\downarrow$& \bf CLAP(\%)$\uparrow$& \bf FAD $\downarrow$& \bf KL $\downarrow$& \bf CLAP(\%)$\uparrow$\\
      \midrule
    1& 2.89& 1.82& 19.8& 4.97& 2.01& 16.0& 5.97& 2.20& 13.2& 6.72& 2.27&12.8\\
 0.8& 2.97& 1.84& 19.6& 4.98& 1.99& 15.8& 5.91& 2.21& 13.2& 6.71& 2.25&12.7\\
 0.6& 3.08& 1.81& 19.5& 4.97& 1.98& 15.2& 5.94& 2.20& 13.6& 6.72& 2.25&12.3\\
      0.4&  3.20& 1.85& 19.5& 5.00& 1.99& 15.6& 6.51&   2.24& 13.6& 6.79& 2.27&12.1\\
      0.2&  3.88& 1.98& 18.3& 5.13&  1.99& 15.5& 6.56&  2.22& 13.1& 6.85& 2.26&11.9\\
      \bottomrule
    \end{tabular}
    }
    \caption{Evaluation of performance metrics across various audio lengths at different cross-frequency compensation percentages ($r$), without same-frequency focusing. Higher values of $r$ imply a lower sparsity rate in the attention operation.}
    \vspace{-2em}
    \label{tbl:result2}
\end{table*}

\subsection{Performance Comparison}
Table~\ref{tbl:main_stable_diffusion} presents a comparative analysis of the experimental results for the original inference, acceleration with \methodname, and acceleration with Token Merging.
It's observed that LiteFocus's acceleration factor increases with the length of the generated audio.
Notably, for 80-second clips, \methodname\ achieves a \efficiency$\times$ faster processing while also improving the audio's quality metrics over those generated by the standard inference. 
Additionally, as the length of the generated audio increases, its three performance metrics—FAD, KL, and CLAP—all deteriorate.
This decline in performance may be attributed to the original model is primarily trained on 10-second audio clips.

When comparing Token Merging's efficiency to \methodname, as shown in Table~\ref{tbl:main_stable_diffusion},we observed that for 80-second audio generation, Token Merging only achieved a 1.56 times faster inference speed, less impressive than \methodname's \efficiency\ times speedup. 
Additionally, unlike LiteFocus, which mitigated the decline in audio quality metrics as audio length increased, audio generated through Token Merging exhibited poorer quality metrics compared to the original inference result.



\subsection{Analysis}

In \methodname, the focus-set $F_i$ associated with each query $Q_i$ is union of distinct sets : one is same-frequency focusing $S_i$, which targets attention within identical frequency bands, and the other part is  cross-frequency compensation $C$ addressing potential gaps missed by $S_i$.



\begin{table}[t]
\centering
    \small
    \resizebox{\linewidth}{!}{
    \begin{tabular}{l  |c c c   }
      \toprule
      \multicolumn{4}{c}{\bf Only Same-Frequency Focusing} \\
      \midrule
      \bf Audio Length& \bf FAD $\downarrow$& \bf KL $\downarrow$& \bf CLAP(\%)$\uparrow$\\
      \midrule
      10 sec& 7.26\ /\ +3.09& 1.98\ /\ +0.12& 16.4\ /\ -1.7\\
      20 sec& 5.10\ /\ +1.38& 1.86\ /\ +0.00& 16.9\ /\ -1.1\\
 30 sec& 5.70 / +0.82& 1.96\ /\ +0.02&15.3\ /\ +0.1\\
      40 sec& 6.06\ /\ +0.66& 2.04\ /\ -0.08& 14.4\ /\ +0.6\\
      80 sec& 6.24\ /\ -0.32& 2.19\ /\ -0.11& 12.5\ /\ +0.3\\

      \bottomrule
    \end{tabular}
    }
    \caption{Performance metrics of \methodname\ with cross-frequency compensation disabled. Each cell presents the metric value (left) and the difference compared to results with cross-frequency compensation enabled (right).}
    \vspace{-2em}
    \label{tbl:result3}
\end{table}

Building upon the original \methodname\ framework, we firstly conducted experiments without the  cross-frequency compensation $C$, and the results are presented in Table~\ref{tbl:result3}. 
We observed that when retaining only same-frequency focusing, the audio generated for durations of 10 seconds and 20 seconds exhibited a decline in quality metrics compared to the original \methodname. 
However, this degradation becomes less pronounced with longer audio. 
Interestingly, for 80-second audio clips, the quality metrics with only same-frequency focusing slightly surpassed those of the original \methodname.


We also conducted \methodname\ experiments exclusively employing cross-frequency compensation, withoud same-frequency focusing, and tested various $r$ percentages across the generation of audio clips of different durations.
Table~\ref{tbl:result2} presents these adjustments' impact on the quality metrics of generated audio clips, illustrating the relationship between reduced cross-frequency compensation ratios $r$ and audio quality.
We can observe that in shorter audio clips (10 seconds and 20 seconds), a lower cross-frequency compensation percentages $r$ leads to a quicker degradation in performance metrics. 
However, this trend slows down in the case of longer audio clips, such as those in 40 and 80 seconds.
This indicates the presence of considerable redundancy in the generation of long audio, suggesting that longer sequences can tolerate a higher degree of sparsity without significant loss in quality.

These findings reveal that same-frequency focusing alone can enhance long audio quality, while optimal cross-frequency compensation improves short audio synthesis. 
This suggests that
attention sparsity based on audio length is key for maintaining high-quality audio generation.

\section{Conclusion and Future Work}
In conclusion, our work presents \methodname, an approach designed to improve the generation of longer audio sequences by latent diffusion models. 
By employing a dual sparse attention mechanism, focusing on same-frequency and cross-frequency compensation, we address some of the computational challenges and performance limitations associated with these models.
This method has shown to reduce inference time and modestly enhance audio quality for longer clips without necessitating model retraining. 
Future work can concentrate on integrating the \methodname\ mechanism directly into the training process to develop inherently efficient models.

\section{Acknowledgement}
This work is supported by the Advanced Research and Technology Innovation Centre (ARTIC), the National University of Singapore under Grant (project number: A0005947-21-00, project reference: ECT-RP2), and the Singapore Ministry of Education Academic Research Fund Tier 1 (WBS: A0009440-01-00).

\newpage
\bibliographystyle{IEEEtran}
\bibliography{main}

\end{document}